\documentclass[10pt,conference]{IEEEtran}
\IEEEoverridecommandlockouts
\usepackage[compress]{cite}
\usepackage[utf8]{inputenc}
\usepackage{amsmath}
\usepackage{amssymb}
\usepackage{mathrsfs}
\usepackage{enumerate}
\usepackage{adjustbox}
\usepackage{xcolor}
\usepackage{graphicx}
\usepackage{algorithm}
\usepackage{algpseudocode}
\usepackage{xpatch}
\usepackage{mathtools}
\usepackage{nicefrac}
\usepackage{float}
\floatstyle{plaintop}
\restylefloat{table}
\usepackage{cleveref}
\DeclarePairedDelimiter{\norm}{\lVert}{\rVert}
\algdef{SE}[DOWHILE]{Do}{doWhile}{\algorithmicdo}[1]{\algorithmicwhile\ #1}%
\algrenewcommand\alglinenumber[1]{{\sffamily\footnotesize#1}}
\makeatletter
\xpatchcmd{\algorithmic}{\itemsep\z@}{\itemsep=.25ex plus2pt}{}{}
\makeatother

\DeclareMathOperator{\blockdiag}{blkdiag}
\DeclareMathOperator*{\argmax}{arg\,max}
\DeclareMathOperator*{\argmin}{arg\,min}
\usepackage{color}
\usepackage{soul}
\usepackage{environ}         
\usepackage{etoolbox}        
\usepackage[acronym]{glossaries}
\newacronym{1g}{1G}{first-generation}
\newacronym{4g}{4G}{fourth-generation}
\newacronym{5g}{5G}{fifth-generation}
\newacronym{mimo}{MIMO}{multiple-input multiple-output}
\newacronym{ris}{RIS}{reconfigurable inteligent surface}
\newacronym{siso}{SISO}{single-input single-output}
\newacronym{mmimo}{mMIMO}{massive multiple-input multiple-output}
\newacronym{cfmmimo}{CF-mMIMO}{cell-free massive multiple-input-multiple-output}
\newacronym{sumimo}{SU-MIMO}{single user MIMO}
\newacronym{mumimo}{MU-MIMO}{multi user MIMO}
\newacronym{embms}{eMBMS}{evolved Multimedia Broadcast and Multicast Service}
\newacronym{sca}{SCA}{successive convex approximation}
\newacronym{sinr}{SINR}{signal-to-interference-plus-noise ratio}
\newacronym{ula}{ULA}{uniform linear array}
\newacronym{uaf}{UatF}{\emph{use-and-then-forget}}
\newacronym{mcs}{MCS}{modulation and coding scheme}
\newacronym{dcc}{DCC}{dynamic cooperation clustering}
\newacronym{mrt}{MRT}{maximum ratio transmission}
\newacronym{ipmmse}{IP-MMSE}{improved partial MMSE}
\newacronym{pzf}{P-ZF}{partial ZF}
\newacronym{zf}{ZF}{zero-forcing}
\newacronym{mr}{MR}{maximum ratio}
\newacronym{se}{SE}{spectral efficiency}
\newacronym{ee}{EE}{energy efficiency}
\newacronym{ap}{AP}{access point}
\newacronym{cpu}{CPU}{central processing unit}
\newacronym{uc}{UC}{user centric}
\newacronym{sse}{SumSE}{sum spectral efficiency}
\newacronym{mise}{MinSE}{minimum spectral efficiency}
\newacronym{asd}{ASD}{angular standard deviation}
\newacronym{adr}{ADR}{aggregated data rate}
\newacronym{embb}{eMBB}{enhanced mobile broadband}
\newacronym{mmtc}{mMTC}{massive machine type communications}
\newacronym{urllc}{URLLC}{ultra reliable low latency communications}
\newacronym{csi}{CSI}{channel state information}
\newacronym{pmi}{PMI}{precoding matrix indicator}
\newacronym{ri}{RI}{rank indicator}
\newacronym{csi-rs}{CSI-RS}{CSI-reference signal}
\newacronym{cri}{CRI}{CSI-RS resource indicator}
\newacronym{bs}{BS}{base station}
\newacronym{re}{RE}{resource element}
\newacronym{mmwave}{mmWave}{millimeter-wave}
\newacronym{umwave}{$\mu$mWaves}{micrometer waves}
\newacronym{rnn}{RNN}{recurrent neural network}
\newacronym{cnn}{CNN}{convolutional neural network}
\newacronym{ngmn}{NGMN}{next-generation mobile network}
\newacronym{lte}{LTE}{Long Term Evolution}
\newacronym{lte-a}{LTE-A}{Long Term Evolution Advanced}
\newacronym{5gnr}{5G NR}{5G New Radio}
\newacronym{mm}{MM}{mixed mode}
\newacronym{cdf}{CDF}{cumulative distribution function}
\newacronym{phy}{PHY}{physical}
\newacronym{mac}{MAC}{medium access control}
\newacronym{3gpp}{3GPP}{3rd Generation Partnership Project}
\newacronym{fdd}{FDD}{frequency division duplexing}
\newacronym{tdd}{TDD}{time division duplexing}
\newacronym{ofdm}{OFDM}{orthogonal frequency division multiplexing}
\newacronym{ss}{SS}{synchronization signal} 
\newacronym{pss}{PSS}{primary synchronization signal} 
\newacronym{sss}{SSS}{secondary synchronization signal} 
\newacronym{pbch}{PBCH}{physical broadcast channel} 
\newacronym{dmrs}{DMRS}{demodulation reference signal} 
\newacronym{gnb}{gNB}{next generation nodeB} 
\newacronym{rsrp}{RSRP}{reference signal received power} 
\newacronym{rrm}{RRM}{radio resource management} 
\newacronym{srs}{SRS}{sounding reference signal} 
\newacronym{ran}{RAN}{radio access network} 
\newacronym{nn}{NN}{neural network} 
\newacronym{ue}{UE}{user equipment} 
\newacronym{awgn}{AWGN}{additive white Gaussian noise} 
\newacronym{epa}{EPA}{Extended Pedestrian A model}
\newacronym{eva}{EVA}{Extended Vehicular A model}
\newacronym{etu}{ETU}{Extended Typical Urban model}
\newacronym{tdl}{TDL}{tapped delay line}
\newacronym{cdl}{CDL}{clustered delay line}
\newacronym{uma}{UMa}{urban macro-cell}
\newacronym{isd}{ISD}{inter-site distance}
\newacronym{nlos}{NLOS}{non-line of sight}
\newacronym{los}{LOS}{line of sight}
\newacronym{o2o}{O2O}{outdoor-to-outdoor}
\newacronym{o2i}{O2I}{outdoor-to-indoor}
\newacronym{ul}{UL}{uplink}
\newacronym{dl}{DL}{downlink}
\newacronym{ls}{LS}{least squares}
\newacronym{mmse}{MMSE}{minimum mean square error}
\newacronym{snr}{SNR}{signal-to-noise ratio}
\newacronym{mse}{MSE}{mean square error}
\newacronym{nr}{NR}{New Radio}
\newacronym{prb}{PRB}{physical resource block}
\newacronym{scs}{SCS}{subcarrier spacing}
\newacronym{bler}{BLER}{block error rate}
\newacronym{smmmra}{SMMMRA}{subgroup multicast \gls{mamimo} resource allocation}
\newacronym{mmf}{MMF}{max-min fairness}
\newacronym{smmu}{SMMU}{subgroups of multicast \gls{mamimo} users}
\newacronym{gsmma}{GSMMA}{greedy subgroup multicast \gls{mamimo} algorithm}
\newacronym{apa}{APA}{adaptive power allocation}
\newacronym{ms}{MS}{mobile station}
\newacronym{cb}{CB}{conjugate beamforming}

\usepackage{flushend}
\newcommand{\herm}{^\mathsf{H}}
\newcommand{\trans}{^\mathsf{T}}
\usepackage{tikz}
\usepackage{changepage}
\usetikzlibrary{fadings}
\tikzfading[name=fade out, 
    inner color=transparent!0,
    outer color=transparent!100]
\usetikzlibrary{patterns}
\usetikzlibrary{shadows.blur}
\usetikzlibrary{shapes}

\ifCLASSOPTIONcompsoc
    \usepackage[caption=false, font=normalsize, labelfont=sf, textfont=sf, subrefformat=parens, labelformat=parens]{subfig}
\else
\usepackage[caption=false, font=footnotesize, subrefformat=parens, labelformat=parens]{subfig}
\fi

\makeatletter
\newcommand\fs@betterruled{%
  \def\@fs@cfont{\bfseries}\let\@fs@capt\floatc@ruled
  \def\@fs@pre{\vspace*{5pt}\hrule height.8pt depth0pt \kern2pt}%
  \def\@fs@post{\kern2pt\hrule\relax}%
  \def\@fs@mid{\kern2pt\hrule\kern2pt}%
  \let\@fs@iftopcapt\iftrue}
\floatstyle{betterruled}
\restylefloat{algorithm}
\makeatother

\begin{document}
\title{User Subgrouping in Scalable Cell-Free\\ Massive MIMO Multicasting Systems
\thanks{This work was supported by the grants PID2020-115323RB-C32 (IRENE-STARMAN), TED2021-131624B-I00 (GERMINAL), TED2021-131975A-I00 (ANTHEM5G), and PID2022-136887NB-I00 (POLIGRAPH) funded by MCIN/AEI/10.13039/501100011033 and, as appropriate by the ``European Union NextGenerationEU/PRTR'', and by the European Union under the Italian National Recovery and Resilience Plan (NRRP) of NextGenerationEU, partnership on ``Telecommunications of the Future'' (PE00000001 - program ``RESTART'', Structural Project 6GWINET, Cascade Call SPARKS, CUP D43C22003080001)}
}

\author{\IEEEauthorblockN{Alejandro de la Fuente\IEEEauthorrefmark{1}, Guillem Femenias\IEEEauthorrefmark{2}, Felip Riera-Palou\IEEEauthorrefmark{2}, Giovanni Interdonato\IEEEauthorrefmark{3}}
\IEEEauthorblockA{\IEEEauthorrefmark{1}Dept. of Signal Theory and Communications, University Rey Juan Carlos, 28942 Fuenlabrada (Madrid), Spain}
\IEEEauthorblockA{\IEEEauthorrefmark{2}Mobile Communications Group, University of the Balearic Islands, 07122 Mallorca (Illes Balears), Spain}
\IEEEauthorblockA{\IEEEauthorrefmark{3}Dept. of Electrical and Information Engineering (DIEI), University of Cassino and Southern Lazio, Cassino, Italy}
Email: alejandro.fuente@urjc.es}

\maketitle

\begin{abstract}
Cell-free massive multiple-input multiple-output (CF-mMIMO) is a breakthrough technology for beyond-5G systems, designed to significantly boost the energy and spectral efficiencies of future mobile networks while ensuring a consistent quality of service for all users. Additionally, multicasting has gained considerable attention recently because physical-layer multicasting offers an efficient method for simultaneously serving multiple users with identical service demands by sharing radio resources.
Typically, multicast services are delivered either via unicast transmissions or a single multicast transmission. This work, however, introduces a novel subgroup-centric multicast CF-mMIMO framework that divides users into several multicast subgroups based on the similarities in their spatial channel characteristics. This approach allows for efficient sharing of the pilot sequences used for channel estimation and the precoding filters used for data transmission. The proposed framework employs two scalable precoding strategies: centralized improved partial MMSE (IP-MMSE) and distributed conjugate beamforming (CB).
Numerical results show that for scenarios where users are uniformly distributed across the service area, unicast transmissions using centralized IP-MMSE precoding are optimal. However, in cases where users are spatially clustered, multicast subgrouping significantly improves the sum spectral efficiency (SE) of the multicast service compared to both unicast and single multicast transmission. Notably, in clustered scenarios, distributed CB precoding outperforms IP-MMSE in terms of per-user SE, making it the best solution for delivering multicast content.
\end{abstract}

\begin{IEEEkeywords}
Cell-free massive MIMO, multicasting, user subgrouping, scalability.
\end{IEEEkeywords}
\glsresetall

\section{Introduction}
\label{sec:introduction}
\Gls{cfmmimo} is an emerging technique for beyond-5G systems owing to its outstanding enhancements in \gls{ee}, \gls{se}, service quality, and reliability \cite{ngo2024ultra}. In a \gls{cfmmimo} system, a large number of \glspl{ap} are distributed across the network and they are connected to a \gls{cpu} via fronthaul links to exchange the \gls{csi} and the user-specific data. \Gls{cfmmimo} exploits the benefits of massive MIMO (mMIMO) and network MIMO, being able to provide \glspl{ms} with nearly uniform service across the coverage area \cite{2017Ngo}. The vast majority of cell-free literature explores the improvements attained in the \gls{se}, \gls{ee}, and coverage performance in \gls{cfmmimo} unicast transmissions. Critically, within the vast volume of data traffic that has been predicted by 2029 \cite{2024Ericsson}, a significant portion will comprise content that can potentially be shared among groups of users in the network, and therefore, can be leveraged through broadcast/multicast techniques \cite{2016delaFuente,2017Araniti}. 
Multicast in mMIMO systems, using both uncorrelated and correlated Rayleigh fading channels, has been proposed and assessed to provide an efficient use of resources \cite{2018SadeghiTWC2,2022delaFuente}. Authors in \cite{2017Doan} first assessed the performance of multicasting in a CF-mMIMO context while proposing a novel \gls{dl} pilot training scheme.
In \cite{2019Zhang}, authors propose a weighted max-min power optimization algorithm that improves the performance when increasing the number of \gls{ap} antennas. 

Despite the benefits revealed by these initial studies on CF-mMIMO multicasting, 
critical issues remain. While precoding and power allocation strategies have been extensively studied in unicast \gls{cfmmimo} scenarios, there is no direct and clear-cut translation between the most efficient configurations for unicast transmissions and those for multicast setups. 
Furthermore, delivering a common data service to a group of MSs has traditionally been accomplished either through a single multicast stream or by sending a unicast stream to each MS in the multicast group. 
Subgroup-centric multicast, as proposed in this research work, can be seen as a mechanism to tailor the transmission of shared content to groups of users experiencing similar intragroup and widely different intergroup propagation conditions \cite{2013Araniti,2018delaFuente,2022delaFuente}. 
Questions regarding the proper management of intragroup and intergroup pilot contamination, the design of common channel estimation processes for all users within the same multicast subgroup, and the impact these common channel estimates might have on the design of the common precoder used to convey the \gls{dl} multicast payload data to users remain largely unsolved.

{\bf Contributions:} 
a novel multicast CF-mMIMO framework is developed considering spatial channel correlation and exploring the advantages and disadvantages of more sophisticated precoding techniques depending on the users' distribution within the \gls{cfmmimo}-multicast framework. 
Users belonging to a multicast group, intended to receive a common service, must share the same uplink (UL) pilot and DL precoder. Our study demonstrates the pilot contamination significance when multicast users experience widely different propagation conditions. As a countermeasure, our work introduces a novel subgroup-centric framework where multicast users are divided into subgroups according to a metric of large-scale propagation similarity and proposes the use of a subgroup-specific pilot and precoder. A comprehensive evaluation of two precoding schemes and power control strategies tailored to the proposed framework is presented, namely, centralized \gls{ipmmse} \cite{2023Femenias} and distributed \gls{cb}, with fractional power control strategies inspired by Demir et al. \cite{2021Demir}. Simulation results demonstrate the benefits of the proposed user subgrouping and power allocation strategies across various system setups, benchmarking the subgroup-centric multicasting approach against conventional multicasting and unicast transmission strategies, considering both uniform and clustered user distributions.


\section{System Model and Multicast Subgrouping Framework}
\label{sec:system_model}
We consider a \gls{cfmmimo} system operating in \gls{tdd} \cite{2017Ngo} that consists of a \gls{cpu} connected via ideal fronthaul links to $L$ \glspl{ap}, each one equipped with $N$ antennas. The \glspl{ap} are deployed over the coverage area to simultaneously provide, either through multicast or unicast, a common data service to $K$ single-antenna \glspl{ms} on the same time-frequency resources. The set of \glspl{ms} is denoted by $\mathcal{K}$ and indexed by $k \!\in\! \mathcal{K}\!=\!\{1,\ldots,K\}$. The set of \glspl{ap} is denoted by $\mathcal{L}$ and indexed by $l \!\in\! \mathcal{L}\!=\!\{1,\ldots,L\}$. 
Each \gls{tdd} frame is divided into \gls{ul} training phase and \gls{dl} payload data transmission phase, whose lengths, measured in samples, are denoted as $\tau_p$ and $\tau_d$, respectively. The frame length is denoted by $\tau_c \!=\! \tau_p \!+\! \tau_d$ and assumed to fit the coherence block.

\subsection{Channel Model}
A conventional block fading channel model is considered wherein the channel is time-invariant and frequency flat within a time-frequency coherence block, and varies independently over different coherence blocks (block fading). The channel response vector $\boldsymbol{h}_{lk} \in \mathbb{C}^{N}$ between the \gls{ap} $l$ and the multicast \gls{ms} $k$, in an arbitrary coherence block\footnote{For the sake of clarity, we omit the index identifying the coherence block.}, is distributed as $\boldsymbol{h}_{lk}  \sim \mathcal{CN}(\boldsymbol{0}_N,\boldsymbol{R}_{lk})$, where $\boldsymbol{R}_{lk} \!\in\! \mathbb{C}^{N \times N}$ is the corresponding positive semi-definite spatial covariance matrix, with average channel gain given by ${\beta}_{lk}  \!=\! \text{tr}\left(\boldsymbol{R}_{lk}\right)/N$.
Assuming that the \glspl{ap} and the \glspl{ms} are well separated, the channel vectors of different \gls{ap}-\gls{ms} pairs can be considered to be independently distributed, i.e., $\mathbb{E}\{\boldsymbol{h}_{l'k'} \boldsymbol{h}_{lk}\herm\} \!=\! \boldsymbol{0}_{N\times N}, \ \forall l'k' \neq lk$. Thus, the channel from \gls{ms} $k$ to the complete set of \glspl{ap} $l \in \mathcal{L}$, 
$\boldsymbol{h}_k\!=\![\boldsymbol{h}_{1k}^T \ldots \boldsymbol{h}_{Lk}^T]^T$, is distributed~as
$\boldsymbol{h}_{k}  \sim \mathcal{CN}(\boldsymbol{0}_{LN},\boldsymbol{R}_{k})$, where $\boldsymbol{R}_{k} = \blockdiag(\boldsymbol{R}_{1k},\ldots,\boldsymbol{R}_{Lk}) \in \mathbb{C}^{LN\times LN}$ is the block-diagonal spatial covariance matrix related to \gls{ms} $k$.

Channel covariance matrices $\boldsymbol{R}_{lk}, \forall~k \in \mathcal{K}, \ \forall~l \in \mathcal{L}$, can be estimated at each \gls{ap} over a large-scale fading time scale (i.e., over multiple coherence blocks) and thus can be safely assumed to be perfectly known at both the \glspl{ap} and the \gls{cpu}~\cite{Bjornson2016a}.

\subsection{Subgroup-Centric \gls{cfmmimo} Multicasting} 
\label{sec:subgrouping}

This work 
proposes delivering a multicast service by subgrouping the $K$ multicast MSs into $G$ disjoint subgroups. In this case, the SE of each multicast subgroup can be adapted to the propagation conditions experienced by the MSs in that particular subgroup, and the multiuser interference reduces to the inter-multicast subgroup interference. We denote the set of multicast subgroups by $\mathcal{G}$ and the subgroups are indexed by $g \in \mathcal{G}=\{1,\ldots,G\}$. The set of \glspl{ms} in subgroup $g$ is denoted by $\mathcal{K}_g$. Letting $K_g=|\mathcal{K}_g|$ be the number of \glspl{ms} in subgroup $g$, it holds that $K = \sum_{g=1}^G K_g$. 
It is interesting to note that in scenarios where the MSs are distributed in spatial clusters, subgrouping the MSs experiencing propagation channels with similar large-scale channel characteristics can bring significant benefits. In this case, MSs in the same multicast subgroup can share the same pilot sequence, thereby controlling pilot contamination. 

In \cite{2018Riera}, the authors benefit from the special attributes of \gls{cfmmimo} channel (distributed transmissions) to propose a K-means-based partitioning algorithm that employs a metric based on the propagation gain vectors $\boldsymbol{\beta}_{k}=[\beta_{1k} \ldots \beta_{Lk}]\trans$ to group \glspl{ms} to improve the pilot allocation when pilot contamination cannot be avoided. 
In the present work, the K-means-based protocol described in \cite{2018Riera} is suitably adapted to classify \glspl{ms} experiencing similar large-scale channel characteristics into multicast subgroups that will benefit from sharing the \gls{ul} pilot sequence to improve both the channel estimation process and the design of the precoding vectors used for \gls{dl} data transmission, improving the SE results.

\begin{figure}[!t]
\centering
\begin{adjustwidth*}{}{-7.5em} 
\resizebox{1.25\columnwidth}{!}{\input{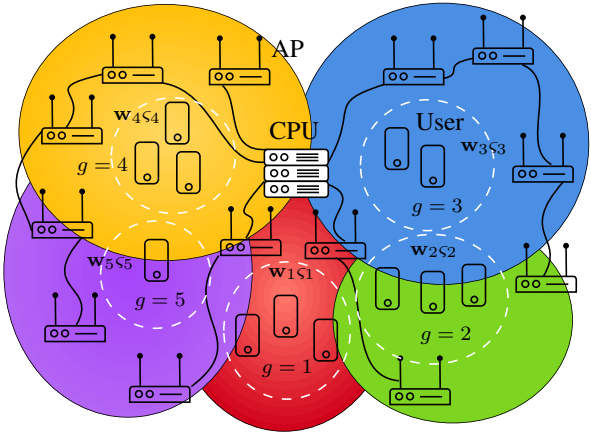}}
\end{adjustwidth*}
\caption{A \gls{cfmmimo} network with multicast user subgrouping (dashed circles) and AP cooperation clustering (blurred colored shapes).}
\label{fig:subgroupingscenario}
\end{figure}

Consequently, a multicast subgroup-centric \gls{cfmmimo} is implemented where the \glspl{ms} in multicast subgroup $g$ are served by a subset of \glspl{ap} somehow readapting the concept of user-centric transmission. We denote the subset of \glspl{ap} serving subgroup $g$ by $\mathcal{L}_g \subseteq \{1,\ldots, L\}$, where $|\mathcal{L}_g|=L_g \leq L$. For later convenience, given a subgroup $g$, we define the set $\mathcal S_g$ as the collection of multicast subgroups served by some (or all) of the \glspl{ap} serving subgroup $g$, that is, $\mathcal{S}_g = \{c: \mathcal{L}_g \cap\mathcal{L}_c \neq \emptyset\}$. The set of multicast subgroups served by \gls{ap} $l$ is denoted as $\mathcal{D}_{l}$. Figure \ref{fig:subgroupingscenario} illustrates the system model of \gls{cfmmimo} multicasting with user subgrouping and \gls{dcc} \cite{2021Demir}.

%

\subsection{Multicast AP Cooperation Clustering and Pilot Allocation}

All \glspl{ms} in the same subgroup are assigned the same pilot sequence chosen among $\tau_p$ available mutually orthogonal pilots (pilot sequence size is equal to $\tau_p$ samples). Note that, as co-pilot \glspl{ms} have linearly dependent channel estimates~\cite{2016Marzetta,2017Bjornsonbook}, the \glspl{ap} cannot separate the \glspl{ms} of the same subgroup in the spatial domain. As a different pilot sequence is assigned to each subgroup $g$, finding the optimal pilot allocation is a combinatorial optimization problem with $\tau_p^G$ possible assignments in a setup with $G$ subgroups and $\tau_p$ pilots; thus, the complexity of evaluating all of them grows exponentially with the number of subgroups. To address this issue, we apply the \gls{dcc} and pilot allocation algorithm proposed by Demir \emph{et al.} in \cite{2021Demir} but suitably adapting it to use multicast subgroup information instead of individual \gls{ms} information. The algorithm will iteratively assign pilots to the subgroups by always selecting the one that leads to the least pilot contamination. To that end, pilots are first assigned to subgroups, and then each \gls{ap} is allowed to serve exactly $\tau_p$ subgroups. For every pilot, the \gls{ap} serves the subgroup with the strongest common average channel gain (i.e., $\frac{1}{K_g} \sum\nolimits_{k \in \mathcal{K}_g} \beta_{lk}$) among the set of subgroups that have been assigned that pilot. 

As shown in Algorithm \ref{alg:pilot_DCC}, the multicast subgroup pilot assignment and cooperation clustering creation process consists of two steps. In the first step, the $\tau_p$ subgroups with indices from $1$ to $\tau_p$ are assigned mutually orthogonal pilots, that is, every \gls{ms} $k \in \mathcal{K}_g$ uses pilot $g$ for $g\in\{1,\ldots,\tau_p\}$. The remaining subgroups, with indices ranging from $\tau_p+1$ to $G$, are then assigned pilots one after the other as follows. Assuming that \gls{ap} $l$ presents an excellent average large-scale propagation gain for \glspl{ms} in subgroup $g$, it is expected to contribute strongly to the service of subgroup $g$ and consequently, it is preferable to assign subgroup $g$ to the pilot for which \gls{ap} $l$ experiences the least pilot contamination. Hence, for each pilot $t$, \gls{ap} $l$ computes the sum of the channel gains $\beta_{lk}$ of the \glspl{ms} that have already been allocated this pilot, and then identifies the index of the pilot minimizing the pilot interference as 
\begin{equation}
    \begin{split}
\tau \leftarrow \argmin\limits_{\boldsymbol{\psi} \in \{1,\ldots,\tau_p\}} \, \sum\limits_{\substack{c \in \mathcal{G},\, c \neq g \\ \boldsymbol{\psi}_g=\boldsymbol{\psi}}} \, \sum\limits_{k \in \mathcal{K}_c} \beta_{lk}.
    \end{split}
\end{equation} 
Pilot $\tau$ is then assigned to subgroup $g$ and the algorithm continues with the next subgroup.

In the second step of the algorithm, the clusters of \glspl{ap} are created as soon as all the subgroups have been assigned to pilots. Each \gls{ap} goes through each pilot and identifies which is the subgroup experiencing the largest common average channel gain among those using that pilot. This subgroup will be served by this particular \gls{ap}. In order to prevent that a subgroup is left unserved, a multicast subgroup $g$ is always served by at least their own Master \gls{ap} $l$, which is selected by the \gls{cpu} according to the rule $l =\argmax \limits_{l \in \mathcal{L}}  \frac{1}{K_g} \sum\nolimits_{k \in \mathcal{K}_g} \beta_{lk}$.

\begin{algorithm}[t!]
\caption{Multicast subgroup pilot assignment and cooperation clustering}\label{alg:pilot_DCC}
\begin{algorithmic}
  \small	
    \State {\underline{\textbf{Input}:} \ $\tau_p$, $\beta_{lk}, G, \mathcal{K}_g$}
    \For{$g=1,\ldots,\tau_p$}
        \State {$\boldsymbol{\psi}_g \leftarrow g$} 
    \EndFor
    \For{$g=\tau_p+1,\ldots,\ G$}
        \State {$l \leftarrow \argmax \nolimits_{l \in \mathcal{L}} \frac{1}{K_g} \sum\nolimits_{k \in \mathcal{K}_g} \beta_{lk}$} 
        
        \State {$\tau \leftarrow \argmin\limits_{\boldsymbol{\psi} \in \{1,\ldots,\tau_p\}}\; \sum\limits_{\substack{c \in \mathcal{G}, \, c \neq g \\ \boldsymbol{\psi}_g=\boldsymbol{\psi}}} \; \sum\limits_{k \in \mathcal{K}_c} \beta_{lk}$} 
        \State {$\boldsymbol{\psi}_g \leftarrow \tau$} 
    \EndFor
    
    \For{$l=1,\ldots,L$}
        \For{$\boldsymbol{\psi}=1,\ldots,\tau_p$}
            \State {$c \leftarrow \argmax \limits_{g \in \{1,\ldots,G\}:\boldsymbol{\psi}_g=\boldsymbol{\psi}} \frac{1}{K_g}\sum\nolimits_{k \in \mathcal{K}_g} \beta_{lk}$} 
             \State {$\mathcal{L}_c \leftarrow \mathcal{L}_c \cup \{l\}$}
        \EndFor
    \EndFor

    \For{$g=1,\ldots,G$}
        
        \If {$\mathcal{L}_g = \{\emptyset\}$}
            \State {$l \leftarrow \argmax \limits_{l \in \mathcal{L}} \frac{1}{K_g} \sum\nolimits_{k \in \mathcal{K}_g} \beta_{lk}$} 
            \State {$\mathcal{L}_g \leftarrow \mathcal{L}_g \cup \{l\}$}
        \EndIf
    \EndFor
   
    \State {\underline{\textbf{Output}:} Pilot assignment $\boldsymbol{\psi}_1,\ldots,\boldsymbol{\psi}_G$ and DCCs $\mathcal{L}_1,\ldots,\mathcal{L}_G$}
\end{algorithmic}
\end{algorithm}

\subsection{Uplink Channel Estimation}

Let $\boldsymbol{\boldsymbol{\psi}}_g \in \mathbb{C}^{\tau_p \times 1}$ be the pilot sequence assigned to subgroup $g$, with $\norm{\boldsymbol{\boldsymbol{\psi}}_g}_2^2=1$. Ideally, pilot sequences should be mutually orthogonal, nonetheless, in practical scenarios it holds that $G>\tau_p$, and a given pilot sequence may be assigned to more than one subgroup, thus resulting in the pilot contamination phenomenon. 
The $N \times \tau_p$ \gls{ul} received pilot signal matrix at \gls{ap} $l$ is
\begin{gather}
    \boldsymbol{Y}_l = \sqrt{\tau_pP_p}\sum^G_{g=1}\sum_{k \in \mathcal{K}_g} \boldsymbol{h}_{lk}\boldsymbol{\boldsymbol{\psi}}\trans_g + \boldsymbol{N}_l,
\end{gather}
where $P_p$ is the per pilot-symbol transmit power of every \gls{ms} and $\boldsymbol{N}_l \in \mathbb{C}^{N \times \tau_p}$ is the \gls{awgn} matrix at \gls{ap} $l$ with i.i.d. elements distributed as $\mathcal{CN}(0,\sigma_u^2)$. To estimate the channel of \glspl{ms} in subgroup $g$, the \gls{ap} projects the received \gls{ul} training signal on the corresponding complex conjugate of the pilot sequence $\boldsymbol{\boldsymbol{\psi}}_g^{\ast}$ to obtain
\begin{equation}
\boldsymbol{y}_{l}^{g} \!=\! \sqrt{\tau_pP_p}\sum_{k \in \mathcal{K}_g} \boldsymbol{h}_{lk}\!+\!\sqrt{\tau_pP_p} \sum_{\substack{c\neq g \\ \boldsymbol{\boldsymbol{\psi}}_c=\boldsymbol{\boldsymbol{\psi}}_g}}\sum_{i\in\mathcal{K}_c} \boldsymbol{h}_{li} \!+ \boldsymbol{n}_{lg},
\end{equation}
where $\boldsymbol{n}_{lg} = \boldsymbol{N}_l \boldsymbol{\boldsymbol{\psi}}^{\ast}_g  \sim \mathcal{CN}(0,\sigma_u^2\mathbf{I}_N)$.

Since the \glspl{ap} cannot separate co-pilot \glspl{ms} in the spatial domain as their channel estimates are correlated, we define the composite channel of the $k \in \mathcal{K}_g$ \glspl{ms} as 
\begin{gather}
\boldsymbol{h}_l^g = \frac{\sqrt{\tau_pP_p}}{K_g}\sum\nolimits_{k \in \mathcal{K}_g}\boldsymbol{h}_{lk},
\label{eq:h_gl}
\end{gather}
which is distributed as $\boldsymbol{h}_l^g  \sim \mathcal{CN}(\boldsymbol{0}_N, \boldsymbol{R}_{l}^g)$, where
\begin{gather}
    \boldsymbol{R}_{l}^g \!=\! \frac{\tau_pP_p}{K^2_g}\sum\nolimits_{k \in \mathcal{K}_g}\boldsymbol{R}_{lk}.
\label{eq:R_estimate}
\end{gather}

The \gls{mmse} channel estimate of the composite channel $\boldsymbol{h}_{l}^g$ can be obtained either at the $l$th \gls{ap} or at the central \gls{cpu} (in cases where the \glspl{ap} are not equipped with local baseband processors) \cite[Sec. 3.2]{2017Bjornsonbook}
\begin{gather}
     \hat{\boldsymbol{h}}_{l}^g = K_g \ \boldsymbol{R}_l^g \ \boldsymbol{\Gamma}_g^{-1} \ \boldsymbol{y}_{l}^g \, , \label{eq:h_gl_est}
\end{gather}
where
\begin{gather}
     \boldsymbol{\Gamma}_g =\tau_p P_p\! \sum\limits_{\substack{\forall c \in \mathcal{G} \\ \boldsymbol{\boldsymbol{\psi}}_c=\boldsymbol{\boldsymbol{\psi}}_g}}\sum\limits_{i\in\mathcal{K}_c} \boldsymbol{R}_{li} \!\! + \!\! \sigma_u^2\boldsymbol{I}_{N}.
\end{gather}
Note that the composite channel estimate is distributed as $\hat{\boldsymbol{h}}_l^g  \sim \mathcal{CN}(\boldsymbol{0}_N, K_g^2 \ \boldsymbol{R}_{l}^g \ \boldsymbol{\Gamma}_g^{-1} \ \boldsymbol{R}_{l}^g)$ and the composite channel estimation error as $\tilde{\boldsymbol{h}}_{l}^g \sim \mathcal{CN}(\boldsymbol{0}_N, \boldsymbol{R}_{l}^g -  K_g^2 \ \boldsymbol{R}_{l}^g \ \boldsymbol{\Gamma}_g^{-1} \ \boldsymbol{R}_{l}^g)$.

\subsection{Downlink Data Transmission and Spectral Efficiency}

Multicast subgrouping performs a multicast transmission by employing unique pre-processing schemes and precoding vectors per subgroup. That is, care is taken to pre-process the common information signal into $G$ uncorrelated signals, each directed to the corresponding multicast subgroup. The received \gls{dl} signal at \gls{ms} $k$ is
\begin{equation}
\begin{split}
      y_{k} \!&=\!  \sum_{l=1}^L\boldsymbol{h}\herm_{lk}  \boldsymbol{x}_{lg} + \sum_{l=1}^L\sum^G_{\substack{c = 1 \\ c \neq g}} \boldsymbol{h}\herm_{lk}  \boldsymbol{x}_{lc} + n_k,   \label{eq:y_gk}
\end{split}
\end{equation}    
where $n_k \sim \mathcal{CN}(0,\sigma_d^2)$ is the \gls{awgn} at \gls{ms} $k$ and $\boldsymbol{x}_{lg}=\boldsymbol{D}_{lg}\mathbf{w}_{lg}\varsigma_g$, with $\mathbf{w}_{lg} \in \mathbb{C}^{N}$ representing the precoding vector used by \gls{ap} $l$ towards multicast subgroup $g$, $\varsigma_g$ denotes the data symbol intended for \glspl{ms} in subgroup $g$, with $\mathbb{E}\{|\varsigma_g|^2\}= 1$, and $\mathbb{E}\{\varsigma_g \varsigma_c^{\ast}\}=0$, $\forall~g \neq c$, and $n_k \sim \mathcal{CN}(0,\sigma_d^2)$ is the \gls{awgn} at \gls{ms} $k$. The set of diagonal matrices  $\boldsymbol{D}_{lg} \in \mathbb{C}^{N\times N}$ are used to describe which \glspl{ap} communicate with which multicast subgroups \cite{2017Buzzi}, and are given by $\boldsymbol{D}_{lg} \!=\! \boldsymbol{I}_N$ if $l \in \mathcal{L}_{g}$, or $\boldsymbol{D}_{lg} \!=\!\boldsymbol{0}_{N\times N}$, otherwise.
Note that, assuming that \gls{ms} $k$ belongs to multicast subgroup $g$, the first term in \eqref{eq:y_gk} denotes the desired signal, whereas the second term is the inter-subgroup interference. 

As we do not transmit \gls{dl} pilots, the achievable \gls{se} of \gls{ms} $k \in \mathcal{K}_g$ must be obtained assuming that it only knows the average value of its effective \gls{dl} channel, that is,  
$\sum_{l=1}^L{\mathbb{E}\left\{\boldsymbol{h}\herm_{lk} \boldsymbol{D}_{lg} \mathbf{w}_{lg}\right\}}$ of \gls{ms} $k \in \mathcal{K}_g$. This is a deterministic number that, thanks to \emph{channel hardening}, can be easily obtained in practice \cite[Theorem 6.1]{2021Demir}, and the corresponding achievable \gls{se} can then be obtained as
\begin{equation}
    \begin{split}
           \text{SE}_{k} = \left(1 - \tau_p/\tau_c\right) \text{log}_2\left(1 + \gamma_{k}\right),
     \end{split}
    \label{eq:SE_gk}
\end{equation}
where $\gamma_{k}$ is the effective \gls{sinr} given by
\begin{equation}
        \!\gamma_{k} \!=\! \frac{\left|\sum\limits_{l=1}^L{\mathbb{E}\left\{\boldsymbol{h}\herm_{lk} \boldsymbol{D}_{lg} \mathbf{w}_{lg}\right\}}\right|^2}{\sum\limits_{c=1}^G \mathbb{E}\left\{\left|\sum\limits_{l=1}^L{\boldsymbol{h}\herm_{lk} \boldsymbol{D}_{lc} \mathbf{w}_{lc}}\right|^2\right\} \!-\! \left|\sum\limits_{l=1}^L{\mathbb{E}\left\{\boldsymbol{h}\herm_{lk} \boldsymbol{D}_{lg} \mathbf{w}_{lg}\right\}}\right|^2 \!\!\!+\! \sigma_d^2},
    \label{eq:SINR_gk}
\end{equation}
and the expectations are with respect to the channel realizations and the channel estimates upon which the precoding vectors are designed.

This is an achievable \gls{se} for \gls{ms} $k$ in subgroup $g$ and holds for any precoding strategy and any multicast \gls{dcc} approach. Nevertheless, we stress that a subgroup is served by a single multicast transmission, which determines a shared \gls{dl} \gls{se} for all the \glspl{ms} in the subgroup. For the data to be able to be reliably decoded by all the \glspl{ms} in the subgroup, the achievable \gls{se} of the multicast subgroup (and thus all \glspl{ms} in this subgroup) will be that of the \gls{ms} experiencing the worst channel conditions in the subgroup. That is, the effective \gls{dl} \gls{se} achievable by all the \glspl{ms} in subgroup $g$ is
\begin{equation}
    \begin{split}
           \text{SE}_{g} = \underset{k \in \mathcal{K}_g}{\text{min}} \text{SE}_{k}.
     \end{split}
    \label{eq:SE_g}
\end{equation}

\subsection{Downlink Precoding and Power Allocation}

We consider two scalable precoding strategies for \gls{cfmmimo}: a centralized \gls{ipmmse} precoding strategy \cite{2023Femenias} and a distributed \gls{cb} precoder \cite{2017Ngo}. Both schemes consider that only the \glspl{ap} in $\mathcal{L}_g$ are computing estimates of the channels for \glspl{ms} in subgroup $g$ and/or send their pilot signals to the \gls{cpu}.

\subsubsection{Centralized IP-MMSE precoding}
We define a composite channel estimate vector per subgroup as $\check{\boldsymbol{h}}^{g} \!=\! \boldsymbol{D}_{g}\hat{\boldsymbol{h}}^{g}$, where $\hat{\boldsymbol{h}}^{g} \!=\! \left[\hat{\boldsymbol{h}}_{1}^{g,\mathsf{T}} \ldots \hat{\boldsymbol{h}}_{L}^{g,\mathsf{T}}\right]\trans$ and $\boldsymbol{D}_{g} \!=\! \blockdiag(\boldsymbol{D}_{1g},\ldots,\boldsymbol{D}_{Lg}) \in \mathbb{C}^{LN\times LN}$. The \gls{cpu} uses the composite channel estimates to compute the composite precoding vectors. 
Since all the \glspl{ms} belonging to subgroup $g$ employ the same pilot, the \gls{cpu} can easily exploit $\check{\boldsymbol{h}}^{g}$ to obtain a scaled vector that points out the direction of the precoding vector. Based on a \emph{virtual} multicast UL-DL duality\footnote{UL-DL duality on the channels defined by \eqref{eq:h_gl} is utilized to design the \emph{virtual} combiners.}, the \gls{ipmmse} combiner \cite{2023Femenias} is given by
\begin{equation*}
           \bar{\mathbf{w}}_g \!=\! \sqrt{\frac{p_g}{\tau_p P_p}}K_g\Bigg(\sum_{c \in \mathcal{S}_{g}} \frac{p_c K_c^2}{\tau_p P_p} \boldsymbol{\check{h}}^{c} \boldsymbol{\big(\check{h}}^{c}\big){\herm} \!+\!  \boldsymbol{Z}_{\mathcal{S}_g} \!+\! \sigma_u^2 \boldsymbol{I}_{L_gN} \Bigg)^{-1} \boldsymbol{\check{h}}^g,
\end{equation*}
where $p_g$ denotes the total amount of power that would be allocated to \glspl{ms} in subgroup $g$ in a \emph{virtual} UL payload transmission phase and 
\begin{equation*}
\boldsymbol{Z}_{\mathcal{S}_g} = \sum\limits_{c \in \mathcal{S}_g} \frac{p_c}{\tau_p P_p}K_c^2 \boldsymbol{D}_g \boldsymbol{\tilde{R}}^c \boldsymbol{D}_g \!+\! \sum\limits_{d \notin \mathcal{S}_g} \frac{p_d}{\tau_p P_p}K_d^2 \boldsymbol{D}_g \boldsymbol{{R}}^d \boldsymbol{D}_g,
\end{equation*}
where $\boldsymbol{\tilde{R}}^c = \blockdiag(\boldsymbol{\tilde{R}}_{1}^c,\ldots,\boldsymbol{\tilde{R}}_{L}^c) \in \mathbb{C}^{LN\times LN}$ denotes the error correlation matrix of the composite channel $\boldsymbol{h}_c$ and $\boldsymbol{{R}}^d = \blockdiag(\boldsymbol{R}_{1}^d,\ldots,\boldsymbol{R}_{L}^d) \in \mathbb{C}^{LN\times LN}$ the covariance matrix of the composite channels of the interfering groups.

The centralized precoding vector used to multicast data to \glspl{ms} in subgroup $g$ can be expressed as
\begin{equation}
    \begin{split}
           \mathbf{w}_g = \sqrt{\rho_g}\frac{\bar{\mathbf{w}}_g}{\sqrt{\mathbb{E}\{\norm{\bar{\mathbf{w}}_g}^2\}}},
     \end{split}
    \label{eq:w_g}
\end{equation}
where $\rho_g \geq 0$ denotes the \gls{dl} transmit power allocated to subgroup $g$. Note that the normalization in (\ref{eq:w_g}) guarantees that $\mathbb{E}\{\norm{\mathbf{w}_g}^2\} = \rho_g$ \cite{2021Demir}.

We propose an \textit{inter-subgroup} fractional DL power control to select the power allocation coefficients proportionally to the trace of $\boldsymbol{R}_l^g$. To satisfy the power constraint at each \gls{ap}, the power allocated to subgroup $g$ is obtained as
\begin{equation}
    \begin{split}
     \rho_{g} = P_{\mathrm{dl}}\frac{\bigg[\sum\limits_{l \in \mathcal{L}_g} \text{tr}\big(\boldsymbol{R}_l^g\big) \bigg]^\nu \omega_g^{-\kappa}}{\underset{\ell \in \mathcal{L}_g}\max \sum\limits_{c \in \mathcal{D}_\ell} \bigg[\sum\limits_{l \in \mathcal{L}_c} \text{tr}\big(\boldsymbol{R}_l^g\big) \bigg]^\nu  \omega_c^{1-\kappa}},
     \end{split}
    \label{eq:p_g}
\end{equation}
where $P_{dl}$ is the total amount of power an \gls{ap} can transmit, $\nu \in [-1,1]$ is the parameter tuning the power allocation in accordance to different policies, i.e., $\nu < 0$ aims at \gls{mmf} characteristics. We also define the largest fraction of $\rho_g$ that can be used at any of the serving \glspl{ap} as
\begin{equation}
    \begin{split}
      \omega_{g} = \underset{l \in \mathcal{L}_g}\max~\mathbb{E}\{\norm{\bar{\mathbf{w}}_{lg}}^2\},
     \end{split}
    \label{eq:omega_g}
\end{equation}
which is used as an additional tuning parameter with an exponent $0 \leq \kappa \leq 1$ that reshapes the ratio of power allocation between different subgroups, where 
$\bar{\mathbf{w}}_g
= \big[\bar{\mathbf{w}}_{1g} \ldots\bar{\mathbf{w}}_{Lg}\big]$.

\subsubsection{Distributed CB precoding}
Distributed operation offers the benefit of deploying new \glspl{ap} without having to upgrade the computational power of the \gls{cpu} since each \gls{ap} contains a local processor that can perform its associated baseband processing tasks and locally designs its transmitted signal \cite{2021Demir}. The distributed \gls{cb} precoding is given by
\begin{equation}
           \mathbf{w}_{lg} = \sqrt{\rho_{lg}}\frac{\boldsymbol{D}_{lg} \boldsymbol{\hat{h}}_{l}^g}{\sqrt{\mathbb{E}\{\norm{\boldsymbol{D}_{lg} \boldsymbol{\hat{h}}_{l}^g}^2\}}}\, ,
\end{equation}
with $\mathbb{E}\{\norm{\boldsymbol{D}_{lg} \boldsymbol{\hat{h}}_{l}^g}^2\} \!=\!K_g^2 \text{tr}(\boldsymbol{D}_{lg}\boldsymbol{R}_l^g\boldsymbol{\Gamma}_g^{-1}\boldsymbol{R}_l^g\boldsymbol{D}_{lg})$.
To satisfy the per-\gls{ap} power constraint, we consider the following \gls{apa} policy~\cite{2019Interdonato}
\begin{align} \label{conn_matrix}
{\rho}_{lg}=\begin{cases}
               P_{dl}\frac{\Big[\text{tr}\big(\boldsymbol{R}_l^g\big) \Big]^\nu}{\sum_{g \in \mathcal{D}_{l}}\Big[\text{tr}\big(\boldsymbol{R}_l^g\big) \Big]^\nu}\, , & \text{ if }  g \in \mathcal{D}_{l} \, \\
               0, & \text{otherwise.}
            \end{cases}
\end{align}

\section{Numerical Results}
\label{sec:results}

We consider a \gls{cfmmimo} network with $L\!=\!100$ \glspl{ap}, each one equipped with $N\!=\!4$ antennas, uniformly distributed at random within a square coverage area of side $1000$ m. To approximate a coverage area without boundaries, the nominal area is wrapped-around by $8$ identical neighbor replicas. The path-loss in dB is given by $-30.5 \!-\! 36.7\text{log}_{10}(d)\!+ F$, where $d$ is the 3D distance between the \gls{ms} and the \gls{ap} and $F$ is the shadow fading, whose standard deviation is $4$ dB, and the decorrelation distance is $9$ m \cite{2017Ngo}. We consider a coherence block of $\tau_c\!=\!200$ samples and a maximum pilot length of $\tau_p\!=\!20$ samples. The pilot transmit power per \gls{ms} is $P_p\!=\!100$ mW, while the maximum \gls{dl} power per \gls{ap} is $P_{dl}\!=\!200$ mW. The \emph{virtual} \gls{ul} power per subgroup to build the centralized \gls{ipmmse} precoder \cite{2023Femenias} is $p_g\!=\!100$ mW. The \gls{dl} power control parameters are set targeting user fairness: $\nu=-0.5$ and $\kappa\!=\!0.5$ for centralized \gls{ipmmse} precoding, and $\nu\!=\!0.5$ for distributed \gls{cb} precoding~\cite{2021Demir}. The angular standard deviation in the local scattering model is $15^{\circ}$. We consider a noise power spectral density of $-174$ dBm/Hz, a receiver noise figure of $7$ dB, and an operating bandwidth of $20$ MHz \cite{2021Demir}. Each simulation result has been obtained as the average of $250$ different snapshots of randomly deployed \glspl{ms} and \glspl{ap} with $500$ channel realizations for each snapshot.

\begin{figure*}[!t]
\subfloat[$100$ \glspl{ms} uniformly distributed]
{\includegraphics[width=0.5\textwidth]
{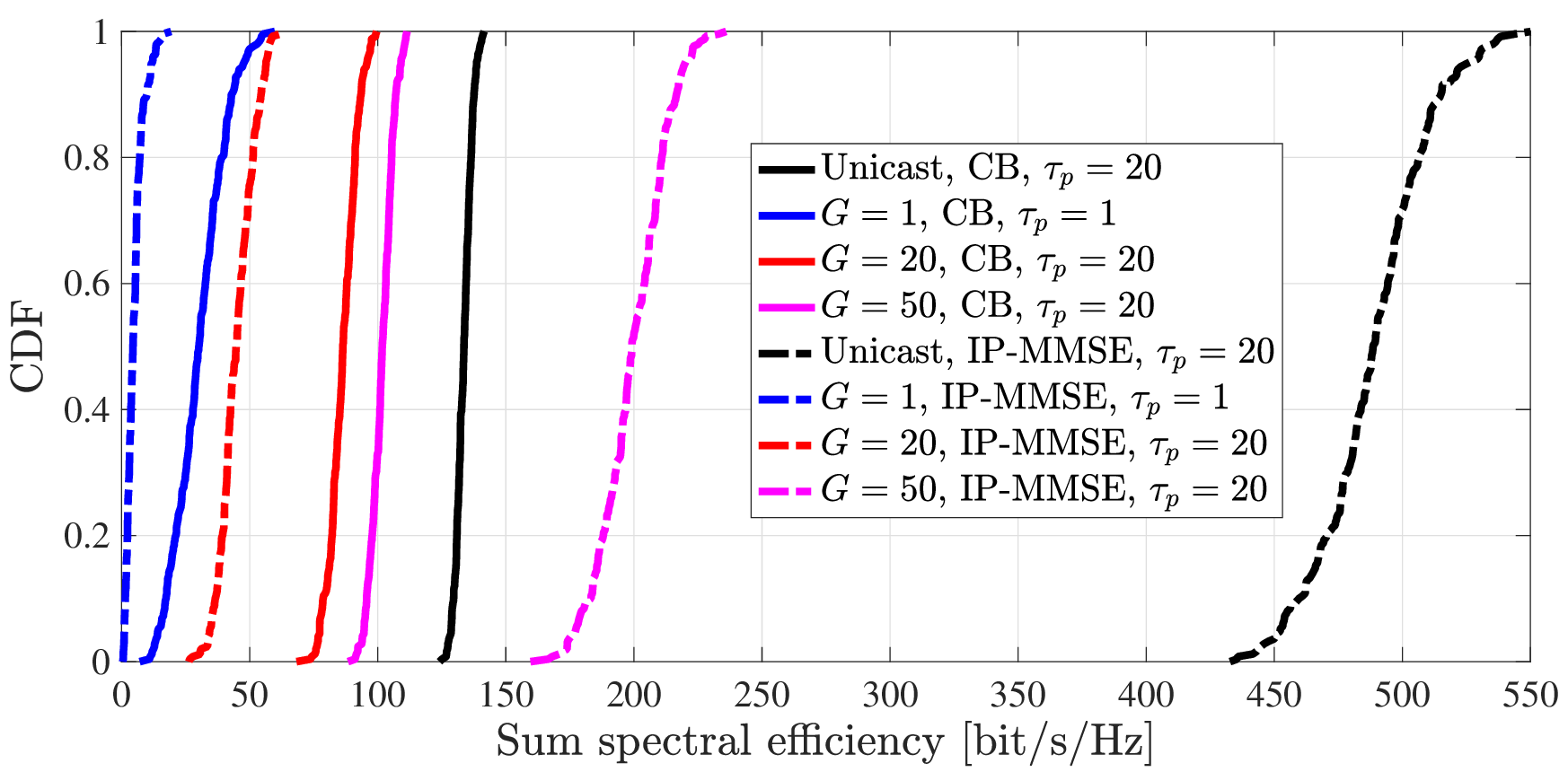}\label{fig:100uniform}}
\subfloat[$500$ \glspl{ms} uniformly distributed]
{\includegraphics[width=0.5\textwidth]
{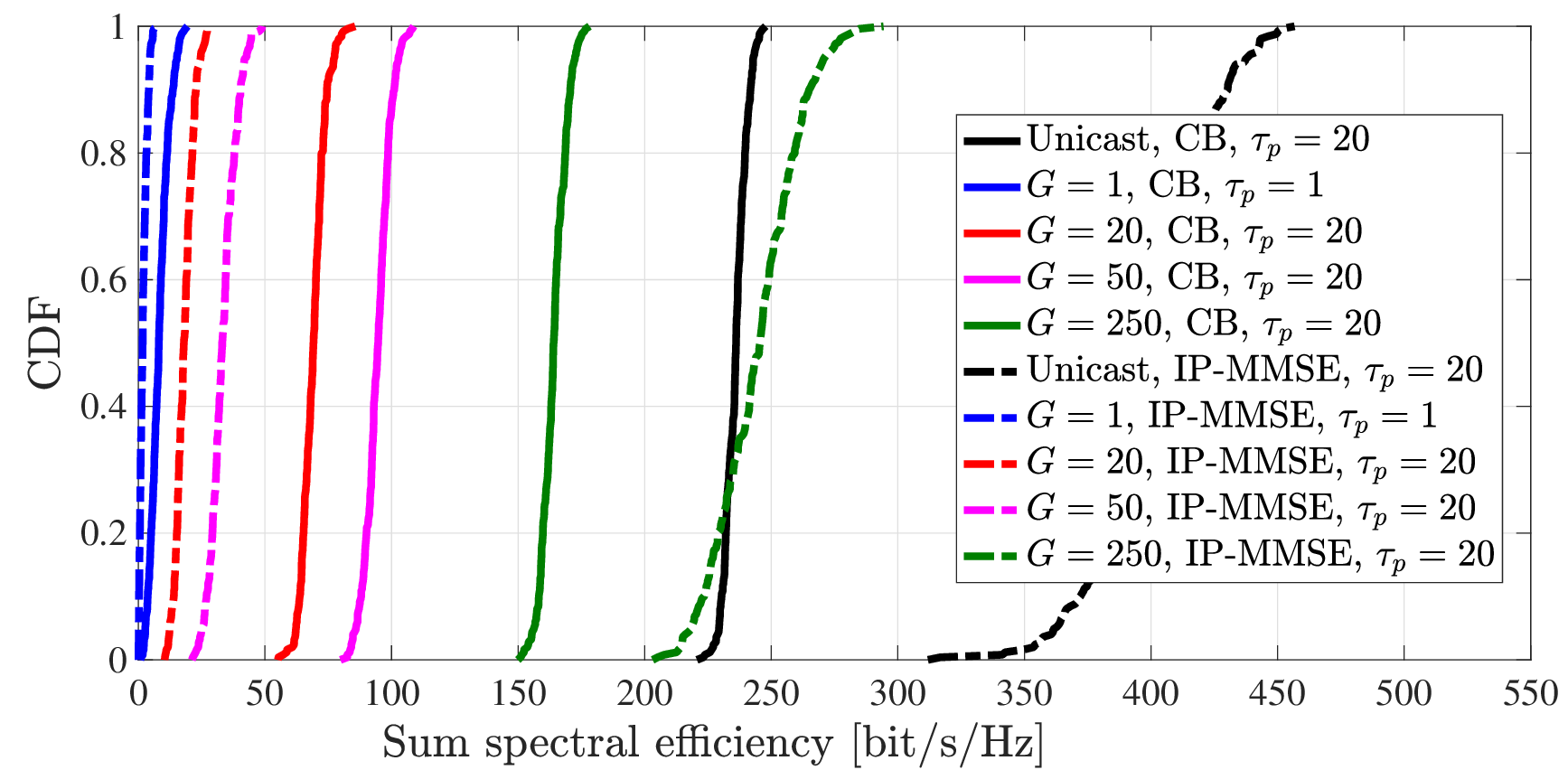}\label{fig:500uniform}}
\caption{{CDF of the sum \gls{se}. Uniform distribution of multicast \glspl{ms}. Unicast vs multicast with CB and IP-MMSE precoding. $L\!=\!100$ APs, $N\!=\!4$}}
\label{fig:uniform}
\subfloat[IP-MMSE precoding]
{\includegraphics[width=0.5\textwidth]{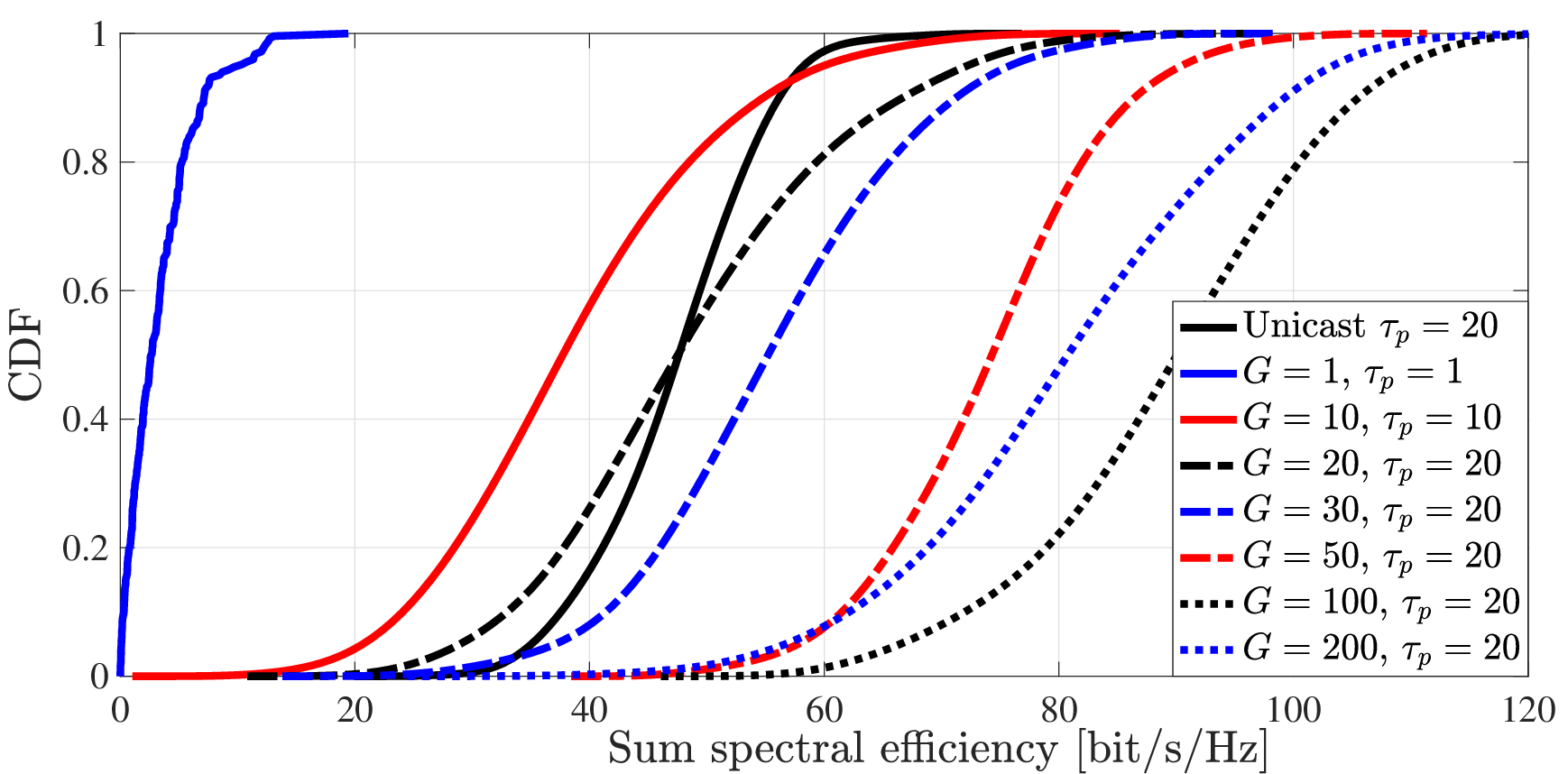}\label{fig:P-MMSE}}
\subfloat[CB precoding]
{\includegraphics[width=0.5\textwidth]{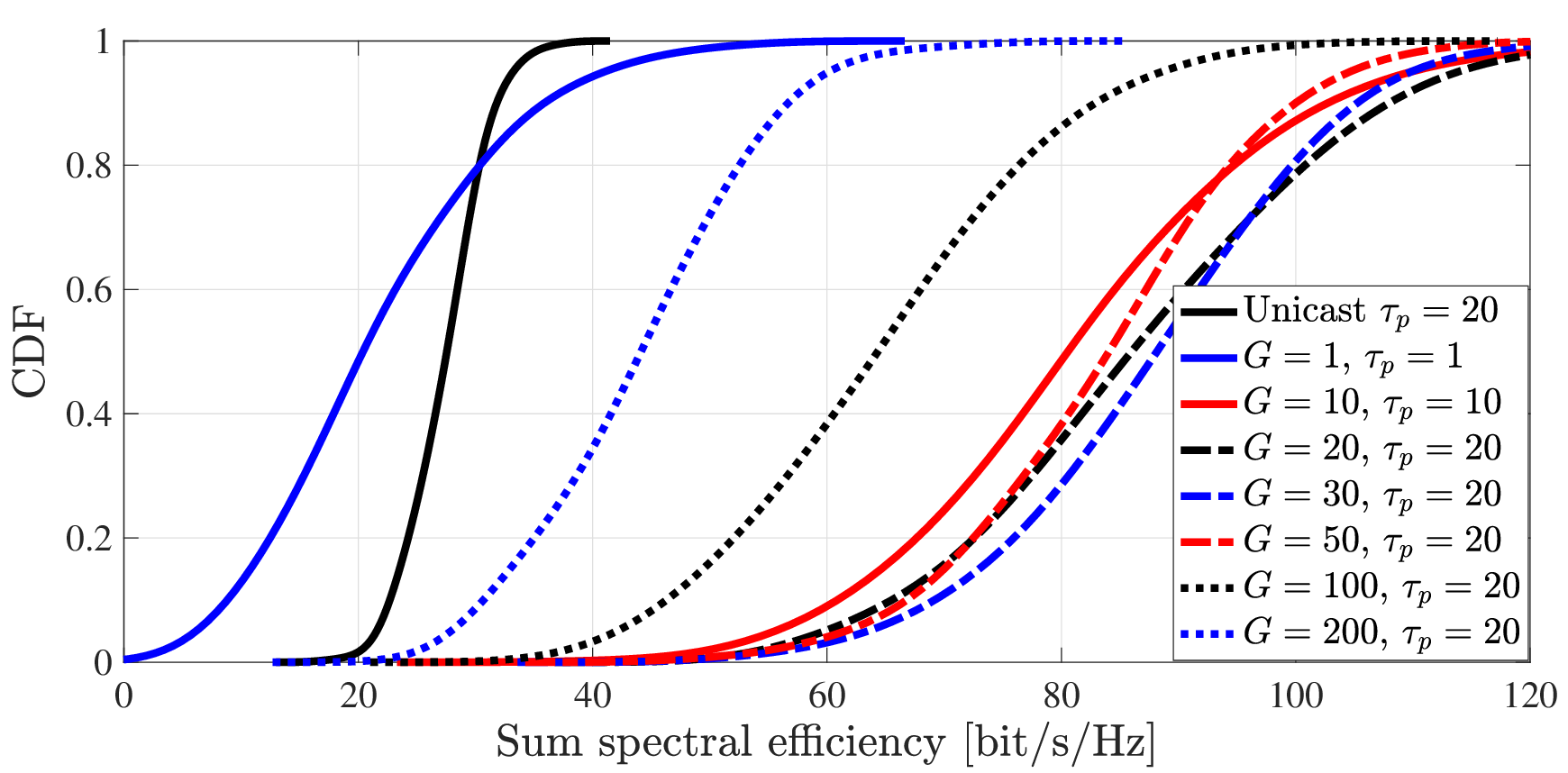}\label{fig:MR}}
\caption{CDF of the sum \gls{se}. Clustered distribution of multicast \glspl{ms} with $10$ spatial clusters of $50$ \glspl{ms}. Unicast vs multicast with CB and IP-MMSE precoding. $L\!=\!100$ APs, $N\!=\!4$.}
\label{fig:10x50all}
\end{figure*}

Figure \ref{fig:uniform} illustrates the \gls{cdf} of the sum \gls{se} achieved by $100$ and $500$ independently and uniformly distributed random multicast \glspl{ms} within the coverage area. We evaluate the performance of either unicast, single multicast or multicast subgrouping transmissions. The results show that when the multicast \glspl{ms} are uniformly distributed, employing unicast transmission, where the channel estimation and the precoding are reasonably accurate despite the pilot contamination (i.e., $\tau_p = 20$ and $K=[100,500]$ in Figs. [\ref{fig:100uniform}, \ref{fig:500uniform}], respectively), performs significantly better than any option employing multicast transmission, under both centralized \gls{ipmmse} and distributed \gls{cb} precoding. Irrespective of the precoder, increasing the number of multicast subgroups improves the performance, but the highest sum \gls{se} is achieved by unicast transmission. 
Furthermore, note that the sum \gls{se} achieved using \gls{ipmmse} is significantly higher than that achieved with \gls{cb}, thus centralized \gls{ipmmse} with unicast transmissions will deliver the highest sum \gls{se} when the multicast \glspl{ms} are uniformly distributed.

To validate the utilization of multicast transmissions, we deploy scenarios where the multicast \glspl{ms} are located in square cluster areas of side $10$ m, thus resulting in bunches of users located very close to one another. This situation can extremely affect the channel estimation and the precoding because of the pilot contamination. Figure \ref{fig:10x50all} shows the \gls{cdf} of the sum \gls{se} of the multicast service when the \glspl{ms} are placed in $10$ spatial clusters of $50$ \glspl{ms} each. 
We observe how the unicast transmissions are severely degraded because of the strong effect of pilot contamination among \glspl{ms} placed in the same spatial cluster (i.e., $50$ \glspl{ms} and $\tau_p=20$ orthogonal pilot sequences). We also notice that using a single multicast transmission does not lead to the best performance, and creating multicast subgroups outperforms both unicast and single multicast transmissions. Fig. \ref{fig:P-MMSE} shows that centralized \gls{ipmmse} precoding, when transmitting to a large number of multicast subgroups, results in the highest sum \gls{se}. Remarkably, this strategy tends to approach unicast transmission while preventing pilot contamination among users located in the same spatial cluster (i.e., $100$ multicast subgroups). \Gls{ipmmse} allows the systems to treat the interference from very close subgroups, and the only reason to not use unicast is the strong pilot contamination from nearby users. In contrast, Fig. \ref{fig:MR} reveals that when using \gls{cb} precoding, splitting the \glspl{ms} into $30$ subgroups presents the best trade-off between signal and interference. 
Indeed, a smaller number of subgroups deteriorates the desired signal due to less accurate channel estimates while a larger number of subgroups increases the interference received from close subgroups.

\section{Conclusion}
\label{sec:conclusions}
This work proposes a novel framework to assess the performance of scalable multicast techniques in \gls{cfmmimo} networks when using different precoders and power allocations. A subgrouping technique has been introduced whereby multicast \glspl{ms} are separated based on their spatial location aiming at improving the performance of the multicast service. Results have shown that unicast transmissions are preferable when the multicast \glspl{ms} are uniformly distributed. 
On the contrary, when the multicast \glspl{ms} tend to form spatial clusters, unicast transmissions are severely degraded by pilot contamination, while multicasting is able to sustain considerable higher rates. Further research will focus on the assessment of heterogeneous deployments where uniform placement of users is combined with the existence of hotspots with larger user density. 


\bibliographystyle{IEEEtran}
\bibliography{IEEEabrv,main}
\end{document}